\documentclass[12pt]{article}
\usepackage{epsfig,amsmath,euscript}

\def\nslash{\rlap{\hspace{0.02cm}/}{n}}

\def\vslash{\rlap{\hspace{0.02cm}/}{v}}

\def\bm#1{\mbox{\boldmath$#1$\unboldmath}}

\begin{document}

\begin{titlepage}

\begin{flushright}
CLNS~04/1898\\
{\tt hep-ph/0411027}\\[0.2cm]
November 1, 2004
\end{flushright}

\vspace{0.7cm}
\begin{center}
\Large\bf 
Impact of Four-Quark Shape Functions on\\ 
Inclusive B Decay Spectra
\end{center}

\vspace{0.8cm}
\begin{center}
{\sc Matthias Neubert}\\
\vspace{0.7cm}
{\sl Institute for High-Energy Phenomenology\\
Newman Laboratory for Elementary-Particle Physics, Cornell University\\
Ithaca, NY 14853, U.S.A.\\[0.3cm]
and\\[0.3cm]
School of Natural Sciences, Institute for Advanced Study\\
Princeton, NJ 08540, U.S.A.}
\end{center}

\vspace{1.0cm}
\begin{abstract}
\vspace{0.2cm}\noindent
It has recently been pointed out that a new class of subleading shape 
functions involving $B$-meson matrix elements of non-local four-quark 
operators contributes at order $\Lambda_{\rm QCD}/m_b$ to 
$\bar B\to X_u\,l^-\bar\nu$ decay distributions in the endpoint region. The 
corresponding functions $f_u(\omega)$ and $f_v(\omega)$ are estimated using 
the vacuum-insertion approximation. A numerical analysis of various 
$\bar B\to X_u\,l^-\bar\nu$ decay spectra suggests that these power 
corrections are very small, below present theoretical uncertainties 
due to other subleading shape-function contributions.
\end{abstract}
\vfil

\end{titlepage}

Inclusive decays of $B$ mesons into final states containing light particles, 
such as $\bar B\to X_u\,l^-\bar\nu$ and $\bar B\to X_s\gamma$, play an 
important role in the extraction of the element $|V_{ub}|$ of the quark 
mixing matrix. Experimental cuts in the analysis of these processes restrict 
the hadronic final state to have large energy, $E_X\sim m_{B}$, but only 
moderate invariant mass, $M_X\sim\sqrt{m_B\Lambda_{\rm QCD}}$. In this region 
of phase space, the inclusive rates can be calculated using a twist 
expansion, which resums infinite sets of local-operator matrix elements into 
non-perturbative shape functions \cite{Neubert:1993ch,Bigi:1993ex}. It is 
well known that the leading term in this expansion obeys a QCD factorization 
formula \cite{Korchemsky:1994jb}, which separates contributions associated 
with the hard scale $m_b$, the jet scale $\sqrt{m_b\Lambda_{\rm QCD}}$, and 
the soft scale $\Lambda_{\rm QCD}$. In recent work, the calculation of 
next-to-leading perturbative corrections to the various components in this 
formula has been completed \cite{Bauer:2003pi,Bosch:2004th}.

Remarkably, for the case of inclusive decay distributions QCD factorization 
can be extended beyond the leading order in the heavy-quark expansion. Using 
the formalism of soft-collinear effective theory (SCET) 
\cite{Bauer:2000yr,Beneke:2002ph}, it has been argued that a factorization 
theorem holds at every order in $\Lambda_{\rm QCD}/m_b$ 
\cite{Lee:2004ja,Bosch:2004cb}. Inclusive decay spectra therefore provide an 
example of a class of observables which have a systematic expansion in 
non-local string operators, built out of quark and gluon fields with 
light-like separation connected by Wilson lines. Non-perturbative hadronic 
physics is encoded in forward $B$-meson matrix elements of these operators in 
heavy-quark effective theory (HQET) \cite{Neubert:1993mb}. The expansion in 
string operators is a generalization of the conventional (local) operator 
product expansion for correlation functions at large (Euclidean) momentum 
transfer.

Recently, two groups have presented the first complete analyses of subleading 
power corrections to arbitrary $\bar B\to X_u\,l^-\bar\nu$ decay 
distributions in the shape-function region \cite{Lee:2004ja,Bosch:2004cb}. In 
these studies, the QCD current-current correlator, whose forward $B$-meson 
matrix element is related to the hadronic tensor $W^{\mu\nu}$, is matched 
onto correlation functions in SCET. These correlators are then expanded in 
terms of non-local string operators in HQET. The resulting expressions 
generalize (and in some cases correct) previous results in the literature 
\cite{Bauer:2001mh,Leibovich:2002ys,Burrell:2003cf}. In particular, it was 
pointed out that there exist some tree-level contributions involving 
subleading shape functions defined in terms of non-local four-quark 
operators, which had not been considered previously. Their contributions to 
arbitrary decay distributions can be parameterized in terms of two functions
$f_u(\omega)$ and $f_v(\omega)$ defined as \cite{Bosch:2004cb}
\begin{equation}\label{fdefs}
   \int d\omega\,e^{-i\omega t}\,\Big[ f_u(\omega)\,T_1
   + f_v(\omega)\,T_4 \Big] 
   = (-i)^2 \int_0^t\!dt_1 \int_{t_1}^t\!dt_2\,
   \frac{\langle\bar B(v)|O_{4q}(t_1,t_2,t)|\bar B(v)\rangle}{2m_B} \,,
\end{equation}
where
\begin{equation}
   O_{4q}(t_1,t_2,t) =
   (\bar h S)_0\,\Gamma_i\,\nslash\,\gamma_\rho^\perp\,t_A\,
   (S^\dagger q)_{t_1 n}\,
   (\bar q S)_{t_2 n}\,\gamma_\perp^\rho\,\nslash\,\Gamma_j\,t_A\,
   (S^\dagger h)_{t n}
\end{equation}
is a non-local string operator with fields ordered along the light-cone 
defined by a vector $n$. Here $h$ is a heavy-quark field in HQET, $q$ is a 
massless quark ($q=u$ for semileptonic decay), $S$ denotes a soft Wilson line, 
and $t_A$ are the generators of color SU$(N_c)$. The light-like vector $n$ 
points in the direction of the final-state hadronic jet $X_u$. We work in the 
$B$-meson rest frame, where $v^\mu=(1,0,0,0)$, and take $n^\mu=(1,0,0,1)$. 
Perpendicular Lorentz indices refer to the transverse plane orthogonal to $v$ 
and $n$ (see \cite{Bosch:2004cb} for further definitions and notation). The 
traces
\begin{equation}
   T_1 = \frac14\,\mbox{tr}\left[ \Gamma_i\,\nslash\,\Gamma_j\,
    \frac{1+\vslash}{2} \right] , \qquad
   T_4 = \frac14\,\mbox{tr}\left[ \Gamma_i\,\nslash\gamma_5\,\Gamma_j\,
    \frac{1+\vslash}{2}\,(\vslash-\nslash)\,\gamma_5 \right]
\end{equation}
depend on the Dirac structures of the flavor-changing weak currents 
$J_i^\dagger=\bar b\,\Gamma_i\,q$ and $J_j=\bar q\,\Gamma_j\,b$ in the 
definition of the hadronic tensor, and $\Gamma_i$, $\Gamma_j$ are in 
principle arbitrary Dirac matrices. Expanding both sides of (\ref{fdefs}) in 
powers of $t$, one finds that the functions $f_u$ and $f_v$ have zero norm 
and first moment, whereas their second moments are given in terms of a local 
four-quark matrix element,
\begin{equation}\label{m2}
   \int d\omega\,\omega^2\,\Big[ f_u(\omega)\,T_1 + f_v(\omega)\,T_4 \Big] 
   = \frac{\langle\bar B(v)|\bar h\,\Gamma_i\,\nslash\,\gamma_\rho^\perp\,
   t_A\,q\,\bar q\,\gamma_\perp^\rho\,\nslash\,\Gamma_j\,t_A\,h
   |\bar B(v)\rangle}{2m_B} \,.
\end{equation}
The subleading shape functions $f_u$ and $f_v$ enter the decay rates 
multiplied by a factor $\pi\alpha_s(\mu_i)\approx 1$, where 
$\mu_i\sim\sqrt{m_b\Lambda_{\rm QCD}}$ is an intermediate matching scale. 
Because of the factor $\pi$, this perturbative coupling does not provide a 
numerical suppression.

While the authors of \cite{Lee:2004ja,Bosch:2004cb} agree on the structural 
form of the four-quark contributions, they differ in the assessment of the 
expected numerical importance of their effects. In \cite{Bosch:2004cb}, it is 
argued that the four-quark contributions are expected to be suppressed with 
regard to other subleading shape functions. On the other hand, the authors of 
\cite{Lee:2004ja} employ arguments based on naive dimensional analysis 
(including counting factors of 4) to speculate that these effects may be the 
dominant source of $\Lambda_{\rm QCD}/m_b$ corrections, which could lead to 
$O(1)$ effects in some decay spectra. In light of this controversy, it may be 
of some value to have well-motivated, if model-dependent estimates of the 
subleading shape functions $f_u$ and $f_v$. This is what we will provide in 
this Letter.

We first consider the case where the flavor of the light quark $q$ matches 
that of the $B$-meson spectator quark. It is empirically well established 
that the magnitude of forward $B$-meson matrix elements of local four-quark 
operators can be estimated by inserting the vacuum intermediate state, 
thereby factorizing them into products of current matrix elements. This 
approximation is routinely used, e.g., in the analysis of lifetime ratios of 
beauty hadrons \cite{Bigi:1994wa,Neubert:1996we}. Depending on the color 
structure of the operators, it is conventional to define 
$\langle\bar B_q|\bar h\,\Gamma_1\,q\,\bar q\,\Gamma_2\,h|\bar B_q\rangle
\equiv {\cal B}_i\,\langle\bar B_q|\bar h\,\Gamma_1\,q|0\rangle\,
\langle 0|\bar q\,\Gamma_2\,h|\bar B_q\rangle$ and
$\langle\bar B_q|\bar h\,\Gamma_1\,t_A\,q\,\bar q\,\Gamma_2\,t_A\,h
|\bar B_q\rangle\equiv\varepsilon_i\,\langle\bar B_q|\bar h\,\Gamma_1\,q
|0\rangle\,\langle 0|\bar q\,\Gamma_2\,h|\bar B_q\rangle$, where the 
subscript ``$i$'' refers to different Dirac structures. The large-$N_c$ 
counting rules of QCD imply ${\cal B}_i=O(1)$ and $\varepsilon_i=O(1/N_c)$, 
and this hierarchy is preserved under renormalization. Theoretical work based 
on lattice QCD \cite{DiPierro:1998ty,Becirevic:2001fy} and QCD sum rules 
\cite{Baek:1998vk} suggests that the $\varepsilon_i$ parameters are indeed 
rather small, typically of order 0.1 or less. (More specifically, 
\cite{Becirevic:2001fy} quotes $\varepsilon_1\approx\varepsilon_2\approx 0.01$ 
at a scale of 2.7\,GeV, whereas \cite{Baek:1998vk} finds 
$\varepsilon_1\approx -0.04$ and $\varepsilon_2\approx 0.06$ at a scale of 
1\,GeV.) This is consistent with empirical findings. For instance, at a 
renormalization point $\mu\approx m_b/2$, the lifetime ratio of charged and 
neutral $B$ mesons can be written as \cite{Neubert:1996we}
\begin{equation}
   \frac{\tau(B^+)}{\tau(B^0)}
   \approx 1 + 0.044 B_1 + 0.003 B_2 - 0.74\varepsilon_1 + 0.20\varepsilon_2
   \stackrel{!}{=} 1.086\pm 0.017 \,,
\end{equation}
where the last result is the current experimental value. If the 
$\varepsilon_i$ parameters were much larger than of order 0.1, a fine tuning 
would be required in order to avoid a large deviation from the experimental 
value. On the other hand, with $B_i\approx 1$ and $\varepsilon_i=O(0.1)$ it is 
easy to reproduce the experimental result.

There appears to be no reason why color suppression should be less effective 
for non-local four-quark operators than for local ones. As a model, we are 
thus led to replace
\begin{eqnarray}\label{model}
   \langle\bar B_q(v)|O_{4q}(t_1,t_2,t)|\bar B_q(v)\rangle
   &\approx& \varepsilon\,
    \langle\bar B_q(v)|\bar h(0)\,[0,t_1]\,\Gamma_i\,\nslash\,
    \gamma_\rho^\perp\,q(t_1 n)|0\rangle \nonumber\\
   &\times & \langle 0|\bar q(t_2 n)\,[t_2,t]\,\gamma_\perp^\rho\,
    \nslash\,\Gamma_j\,h(t n)|\bar B_q(v)\rangle \,,
\end{eqnarray}
where $[t_k,t_l]$ denotes a straight soft Wilson line connecting the points 
$t_k n$ and $t_l n$, and $\varepsilon$ is the color-suppression factor. This 
ansatz completely specifies our model for the subleading shape functions. 
Whereas the definitions of the parameters ${\cal B}_i$ and $\varepsilon_i$ 
introduced in \cite{Neubert:1996we} are completely general, the introduction 
of the parameter $\varepsilon$ in the equation above corresponds to a model 
hypothesis, because we assume that $\varepsilon$ is independent of the 
position arguments $t_1$, $t_2$, $t$ and of the Dirac structures $\Gamma_i$, 
$\Gamma_j$. Nevertheless, as we shall see, our model provides an expression 
for the subleading shape functions $f_u$ and $f_v$ with all the right 
properties such as the correct support and moment relations, as far as they 
are known. We expect that it predicts the rough overall scale of the effect 
reliably. In fact, many analyses of inclusive $B$ decays are based on 
measurements of partial decay rates over kinematical domains that fall in 
between the shape-function region and the region where a conventional 
operator product expansion can be applied \cite{Bosch:2004th}. In such a 
situation, the dominant contributions to the decay rates are associated with 
the lowest non-zero moments of the shape functions. In the case of the 
four-quark shape functions, the lowest non-zero moments are given by the 
matrix element of the local four-quark operator in (\ref{m2}), for which the 
vacuum insertion approximation is certainly reasonable.

The $B\to$~vacuum matrix elements of the non-local quark bilinears in 
(\ref{model}) can be expressed in terms of the leading light-cone distribution 
amplitude of the $B$ meson \cite{Grozin:1996pq},
\begin{equation}
   \langle 0|\bar q(t' n)\,[t',t]\,\nslash\,\Gamma\,h(t n)|\bar B_q(v)\rangle 
   = - \frac{if_B m_B}{2}\,\mbox{tr}\left[ \nslash\,\Gamma\,
   \frac{1+\vslash}{2}\,\gamma_5 \right]
   \int_0^\infty\!d\omega\,\phi_+(\omega)\,
   e^{-i\omega t'-i(\bar\Lambda-\omega)t} ,
\end{equation}
where we have used that the $B$ meson in HQET carries momentum 
$\bar\Lambda v$, with $\bar\Lambda=m_B-m_b$. Throughout this note we work at
lowest order in perturbation theory, so that we can ignore the scale 
dependence of the various objects in the above equation. Taking into account 
that the distribution amplitude $\phi_+$ is real, we obtain in our model
\begin{eqnarray}
   &&\hspace{-0.55cm} (-i)^2 \int_0^t\!dt_1 \int_{t_1}^t\!dt_2\,
   \langle\bar B_q(v)|O_{4q}(t_1,t_2,t)|\bar B_q(v)\rangle \nonumber\\
   &=& \varepsilon\,\frac{f_B^2 m_B^2}{4}\,e^{-i\bar\Lambda t}\,
    \mbox{tr}\left[ \gamma_5\,\frac{1+\vslash}{2}\,\Gamma_i\,\nslash\,
    \gamma_\rho^\perp \right]\cdot
    \mbox{tr}\left[ \gamma_\perp^\rho\,\nslash\,\Gamma_j\,
    \frac{1+\vslash}{2}\,\gamma_5 \right] \nonumber\\
   &\times& \int_0^\infty\!d\omega_1 \int_0^\infty\!d\omega_2\,
    \phi_+(\omega_1)\,\phi_+(\omega_2)\,
    \left[ \frac{1}{\omega_1\omega_2} + \frac{1}{\omega_1-\omega_2} 
    \left( \frac{e^{i\omega_1 t}}{\omega_1}
    - \frac{e^{i\omega_2 t}}{\omega_2} \right) \right] .
\end{eqnarray}
The product of traces can be expressed in terms of the objects $T_1$ and 
$T_4$ defined earlier, using the fact that between 
$\nslash\dots\frac12(1+\vslash)$ any Dirac matrix can be decomposed into the 
basis $\bm{1}$, $\gamma_5$, $\gamma_\perp^\mu$. A straightforward calculation 
shows that
\begin{equation}
   T_1 + T_4 = \frac14\,
   \mbox{tr}\left[ \gamma_5\,\frac{1+\vslash}{2}\,\Gamma_i\,\nslash\,
    \gamma_\rho^\perp \right]\cdot
   \mbox{tr}\left[ \gamma_\perp^\rho\,\nslash\,\Gamma_j\,
    \frac{1+\vslash}{2}\,\gamma_5 \right] .
\end{equation}
Taking the Fourier transform of (\ref{fdefs}), we then obtain
\begin{equation}
   f_u(\omega) = f_v(\omega)
   = - \varepsilon\,\frac{f_B^2 m_B}{2}\,g(\omega) \,,
\end{equation}
where
\begin{equation}
   g(\omega) = \delta(\bar\Lambda-\omega) \left[ \int_0^\infty\!d\omega'\,
   \frac{\phi_+(\omega')}{\omega'} \right]^2
   + \frac{2\phi_+(\bar\Lambda-\omega)}{\bar\Lambda-\omega}\,
   \mbox{P} \int_0^\infty\!d\omega'\,
   \frac{\phi_+(\omega')}{\bar\Lambda-\omega-\omega'} \,.
\end{equation}
``P'' denotes the principal-value prescription. The function $g$ has support 
on the half-interval $-\infty<\omega\le\bar\Lambda$, as is indeed required for 
all shape functions \cite{Neubert:1993ch,Bigi:1993ex}. Its first few moments 
are
\begin{equation}
   \int d\omega\,g(\omega) = 0 \,, \qquad
   \int d\omega\,\omega\,g(\omega) = 0 \,, \qquad
   \int d\omega\,\omega^2 g(\omega) = 1 \,.
\end{equation}
The first two conditions ensure that the subleading shape functions $f_u$ and 
$f_v$ have vanishing norm and first moment, as required on general grounds
\cite{Bosch:2004cb}.

\begin{figure}[t]
\begin{center}
\epsfig{file=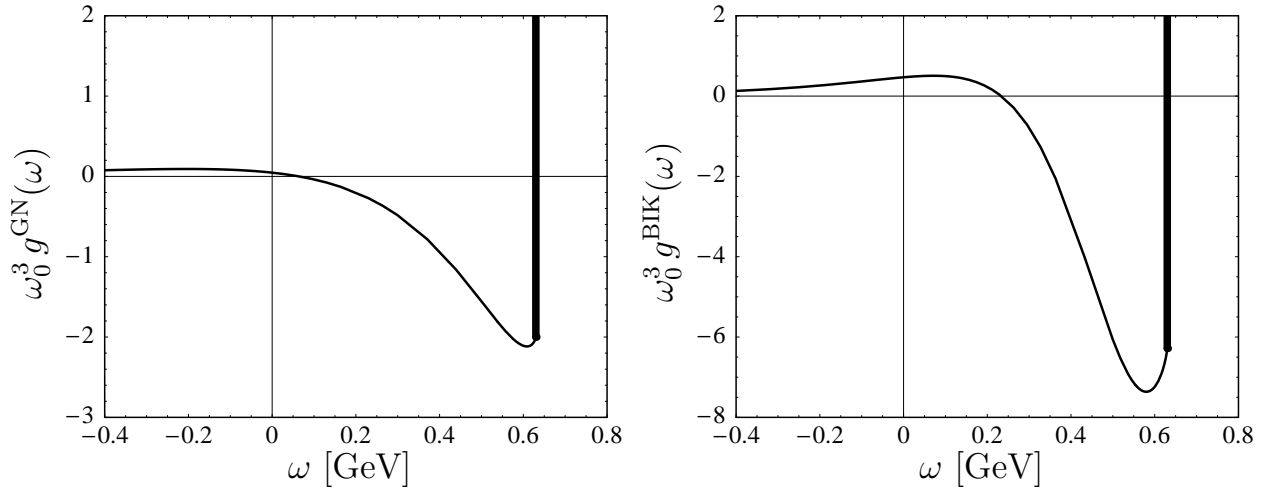,width=\textwidth}
\end{center}
\vspace{-0.2cm}
\centerline{\parbox{14.5cm}{\caption{\label{fig:gfun}
Model predictions for the function $g(\omega)$ obtained using 
$\bar\Lambda=0.63$\,GeV, corresponding to $m_b=4.65$\,GeV. The thick lines 
represent the $\delta$-function localized at $\omega=\bar\Lambda$.}}}
\end{figure}

In order to have some explicit models at hand, we adopt two forms of the  
$B$-meson light-cone distribution amplitude motivated by QCD sum rules
\cite{Grozin:1996pq,Braun:2003wx}, namely
\begin{eqnarray}\label{phiGN}
   \phi_+^{\rm GN}(\omega)
   &=& \frac{\omega}{\omega_0^2}\,e^{-\omega/\omega_0} \,; \hspace{1.08cm}
    \omega_0 = \frac23\,\bar\Lambda \,, \nonumber\\
   \phi_+^{\rm BIK}(\omega)
   &=& \frac{2\omega_0^2\omega}{(\omega^2+\omega_0^2)^2} \,; \qquad
    \omega_0 = \frac{8}{3\pi}\,\bar\Lambda \,,
\end{eqnarray}
where the relation between $\omega_0$ and $\bar\Lambda$ is implied by the 
equations of motion. Introducing the dimensionless variable 
$z=(\bar\Lambda-\omega)/\omega_0=\hat\omega/\omega_0\ge 0$, we obtain
\begin{eqnarray}\label{gGN}
   g^{\rm GN}(\omega)
   &=& \frac{1}{\omega_0^3} \left[
    \delta(z) - 2 e^{-z} + 2z\,e^{-2z}\,\mbox{Ei}(z) \right] , \nonumber\\
   g^{\rm BIK}(\omega)
   &=& \frac{1}{\omega_0^3} \left[
    \frac{\pi^2}{4}\,\delta(z) + \frac{8}{(1+z^2)^4} 
    \left( z\ln z + \frac{z}{2}\,(1+z^2) - \frac{\pi}{4}\,(1-z^2) \right)
    \right] ,
\end{eqnarray}
where $\mbox{Ei}(z)=-\mbox{P}\int_{-z}^\infty dt\,e^{-t}/t$ is the 
exponential-integral function. The variable $\hat\omega=\bar\Lambda-\omega$ 
is the most convenient one when calculating decay spectra, because it is 
independent of the definition used for the $b$-quark mass 
\cite{Bosch:2004th}. A graphical representation of these two functions is 
shown in Figure~\ref{fig:gfun}. Notice the rapid fall-off of the model 
functions for values $\omega$ away from the endpoint. For the first model, 
this reflects the assumed exponential fall-off of the distribution amplitude 
$\phi_+$. But even in the second model, for which $\phi_+$ has only a 
power-like fall-off, the function $g$ decreases quickly in magnitude. 

So far, we have assumed that the light quark $q$ is contracted with the 
spectator quark in the $B$ meson. But this is not always possible (e.g., if 
they have different flavor), and one may entertain the possibility that the 
light-quark pair in the operator $O_{4q}$ is produced from the vacuum. In this 
case, a possible factorization model is to write
\begin{eqnarray}
   &&\hspace{-0.55cm} 
    \langle\bar B(v)|(\bar h S)_0\,\Gamma_1\,t_A\,(S^\dagger q)_{t_1 n}\,
    (\bar q S)_{t_2 n}\,\Gamma_2\,t_A\,(S^\dagger h)_{t n}|\bar B(v)\rangle
    \nonumber\\
   &\approx& {\cal B}\,
    \langle\bar B(v)| \big[ (\bar h S)_0\,\Gamma_1\,t_A \big]_\alpha^i\,
    \big[ \Gamma_2\,t_A\,(S^\dagger h)_{t n} \big]_\beta^j|\bar B(v)\rangle\,
   \langle 0| \big[ (S^\dagger q)_{t_1 n} \big]_\alpha^i\,
    \big[ (\bar q S)_{t_2 n} \big]_\beta^j |0\rangle \,,
\end{eqnarray}
where ${\cal B}$ is a bag parameter, and we have left the Dirac structures 
arbitrary for the moment. Lorentz and gauge invariance dictate that
\begin{equation}
   \langle 0| \big[ (S^\dagger q)_{t_1 n} \big]_\alpha^i\,
   \big[ (\bar q S)_{t_2 n} \big]_\beta^j |0\rangle
   = - \frac{\delta^{ij}}{4N_c}\,\langle\bar q q\rangle\,
   \Big[ \bm{1} + i\Lambda_h (t_2-t_1) \nslash \Big]_{\alpha\beta} \,,
\end{equation}
where $\langle\bar q q\rangle$ is the local quark condensate, and $\Lambda_h$ 
is a hadronic parameter. From above, we now obtain in our model
\begin{eqnarray}
   &&\hspace{-0.55cm} 
    \langle\bar B(v)|(\bar h S)_0\,\Gamma_1\,t_A\,(S^\dagger q)_{t_1 n}\,
    (\bar q S)_{t_2 n}\,\Gamma_2\,t_A\,(S^\dagger h)_{t n}|\bar B(v)\rangle
     \nonumber\\
   &=& - \frac{{\cal B}\,C_F}{4N_c}\,\langle\bar q q\rangle\
    \Big[ \langle\bar B(v)| \bar h(0)\,[0,t]\,\Gamma_1\,\Gamma_2\,h(t n)
    |\bar B(v)\rangle \nonumber\\[-0.2cm]
   &&\hspace{2.4cm}\mbox{}+ i\Lambda_h (t_2-t_1)\,
    \langle\bar B(v)| \bar h(0)\,[0,t]\,\Gamma_1\,\nslash\,\Gamma_2\,h(tn)
    |\bar B(v)\rangle \Big] \,.
\end{eqnarray}
Applying this formula to the specific case of the operator in (\ref{fdefs}), 
for which $\Gamma_1=\Gamma_i\,\nslash\,\gamma_\rho^\perp$ and
$\Gamma_2=\gamma_\perp^\rho\,\nslash\,\Gamma_j$, we see that the Dirac 
structures $\Gamma_1\,\Gamma_2$ and $\Gamma_1\,\nslash\,\Gamma_2$ both vanish 
due to $\nslash^2=0$. Therefore, the non-valence contributions vanish in the 
vacuum-insertion approximation.

Let us briefly discuss the phenomenological implications of our results by
considering three decay spectra in semileptonic $\bar B\to X_u\,l^-\bar\nu$ 
decay, referring to \cite{Bosch:2004cb} for details and derivations. In all 
cases we show the contributions of the leading-order shape function 
$\hat S(\hat\omega)$ and of the four-quark shape functions as calculated in 
our model, ignoring other subleading shape-function contributions, which have 
already been discussed in \cite{Bosch:2004cb}. In our model, only decay 
distributions of charged $B$ mesons are effected by the four-quark 
contributions. Of particular interest for a measurement of $|V_{ub}|$ are the 
spectra in the variables $E_l$ (charged-lepton energy), $P_+$ (hadronic 
energy minus momentum), and $s_H$ (hadronic invariant mass squared). The 
corresponding normalized distributions are given by
\begin{eqnarray}\label{rates}
   \frac{1}{\Gamma}\,\frac{d\Gamma}{dE_l}
   &=& 4 \int\limits_0^{m_B-2E_l}
    \frac{d\hat\omega}{m_B-\hat\omega} \left[ \hat S(\hat\omega)
    + 3\varepsilon\pi\alpha_s f_B^2\,\hat g(\hat\omega) \right] ,
    \nonumber\\
   \frac{1}{\Gamma}\,\frac{d\Gamma}{dP_+}
   &=& \hat S(P_+)
    + \frac23\,\varepsilon\pi\alpha_s f_B^2\,\hat g(P_+) \,, \nonumber\\
   \frac{1}{\Gamma}\,\frac{d\Gamma}{ds_H}
   &=& \frac{1}{m_B} \int\limits_{s_H/m_B}^\infty
    \frac{d\hat\omega}{\hat\omega}\,
    F\!\left( \hat\omega,\frac{s_H}{m_B\hat\omega} \right) ,
\end{eqnarray}
where $\hat g(\hat\omega)=g(\bar\Lambda-\hat\omega)$, and in the latter case
\begin{equation}
   F(\hat\omega,r) = 2r^2(3-2r)\,\hat S(\hat\omega)
   + 4r(1-r)\,\varepsilon\pi\alpha_s f_B^2\,\hat g(\hat\omega) \,.
\end{equation}
Note that the contribution to the charged-lepton spectrum is much larger than
that in the other two cases. We have studied the numerical impact of the 
four-quark contributions to these spectra using the model functions in 
(\ref{gGN}) with $\bar\Lambda=0.63$\,GeV and the default choice for the 
leading-order shape function from \cite{Bosch:2004cb}. We take $f_B=200$\,MeV 
for the $B$-meson decay constant and $\alpha_s=0.3$ for the strong coupling at 
the intermediate scale. Even without assuming a significant color suppression 
the effects are very small. For $\varepsilon=1$, sizable distortions of the 
$P_+$ and $s_H$ spectra occur below $P_+\approx 0.5$\,GeV and 
$s_H\approx 1.5$\,GeV$^2$, while those of the charged-lepton energy spectrum 
are located above $E_l\approx 2.2$\,GeV. For more realistic values 
$|\varepsilon|\ll 1$, the effects are even smaller. Varying the input 
parameters within reasonable limits does not change this conclusion. The 
four-quark contributions are proportional to the combination 
$\varepsilon\pi\alpha_s f_B^2/\bar\Lambda^3$ times a dimensionless function of 
the ratio $\hat\omega/\bar\Lambda$. Assuming $f_B=(200\pm 30)$\,MeV and 
$\bar\Lambda=(0.63\pm 0.07)$\,GeV, we have $\varepsilon\pi\alpha_s f_B^2/%
\bar\Lambda^3=(0.15\pm 0.07)\varepsilon$\,GeV$^{-1}$, where the parametric 
uncertainty can be absorbed into our ignorance about the color-suppression 
factor $\varepsilon$.

\begin{figure}[t]
\begin{center}
\epsfig{file=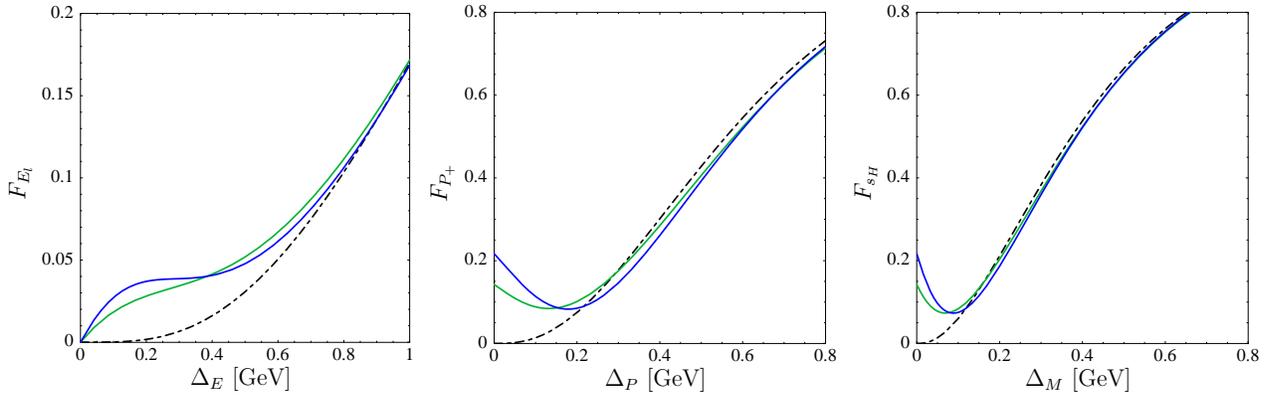,width=\textwidth}
\end{center}
\vspace{-0.2cm}
\centerline{\parbox{14.5cm}{\caption{\label{fig:spectra}
Model predictions for the partial $\bar B\to X_u\,l^-\bar\nu$ rate fractions 
with cuts on $E_l$, $P_+$, and $s_H$, respectively. The dashed line shows the
leading-order result, while the two solid lines include the contributions from 
four-quark shape functions, evaluated setting $\varepsilon=1$.}}}
\end{figure}

Figure~\ref{fig:spectra} shows results for the partial rate fractions 
obtained by integrating the spectra in (\ref{rates}) over 
$E_l\ge E_0=(m_B-\Delta_E)/2$, $P_+\le\Delta_P$, and $s_H\le s_0=m_B\Delta_M$, 
where the quantities $\Delta_i$ are chosen such that in all three cases the
charm background starts at $\Delta_i=m_D^2/m_B\approx 0.66$\,GeV. The effects
of the four-quark shape functions are included under the extreme assumption of 
no color suppression ($\varepsilon=1$). In reality, we expect their effects to
be significantly smaller. If we consider the partial rate fractions obtained 
by integrating the spectra over the phase space not contaminated by charm 
background, we find that the largest effects in the two models lead to 
\begin{equation}
   F_{E_l} = (6.5 + 1.4\varepsilon)\% \,, \qquad
   F_{P_+} = (60.9 - 2.3\varepsilon)\% \,, \qquad
   F_{s_H} = (80.6 - 0.7\varepsilon)\% \,.
\end{equation}
In the latter two cases these corrections are negligible even if the 
color-suppression factor $\varepsilon$ is not particularly small. 

For comparison, we note that the authors of \cite{Lee:2004ja} estimate (using 
naive dimensional analysis, and ignoring color suppression) that the impact of 
four-quark shape functions on the charged-lepton energy fraction could be as 
large as 180\% of the leading term. Even for $\varepsilon=1$ this would be an 
order of magnitude bigger than our estimate. The main reason is that these 
authors include an ``enhancement factor'' $g_s^2=4\pi\alpha_s\approx 4$ in 
their estimate of the power-suppressed effects, as opposed to, say, 
$\pi\alpha_s\approx 1$. Clearly, dimensional analysis cannot tell the 
difference between these two factors. The precise numerical coefficients 
multiplying the four-quark shape-function contributions to a given observable 
can only be determined by explicit calculation, as done in (\ref{rates}). Note
that following the reasoning of \cite{Lee:2004ja} one would associate an
enhancement factor $g_s^2=4\pi\alpha_s\sim 12$ with soft-gluon exchange (for 
which $\alpha_s\sim 1$), which would be wrong, since the corresponding powers 
of $g_s$ are already included in what is conventionally 
called $\Lambda_{\rm QCD}$. 

In summary, we have presented a simple, but well-motivated model for the
four-quark subleading shape functions $f_u(\omega)$ and $f_v(\omega)$, which
contribute at order $\Lambda_{\rm QCD}/m_b$ to inclusive $B$-decay spectra in 
the endpoint region. We have shown that in the vacuum-insertion approximation 
non-zero contributions only arise from four-quark operators for which the 
light quark flavor matches that of the $B$-meson spectator quark. Vacuum pair 
production of the light quark pair, though not suppressed on general grounds, 
does not contribute due to the Dirac structure of the relevant operators. The 
result for the subleading shape functions $f_u$ and $f_v$ is given by a 
color-suppression factor times a double convolution integral over a product of 
two $B$-meson light-cone distribution amplitudes. Using two simple models for 
this function, we have obtained explicit forms for the subleading shape 
functions, which are compatible with all known constraints from analyticity 
and moment relations. The corresponding impact on the decay distributions in 
semileptonic $\bar B\to X_u\,l^-\bar\nu$ decays is found to be negligible. 
While the results reported here are admittedly model dependent, we believe 
that they support our earlier claim \cite{Bosch:2004cb} that these four-quark 
contributions are likely to be smaller than other subleading shape-function 
effects.

{\em Acknowledgments:\/}
I am grateful to Bj\"orn Lange for useful discussions. This research was 
supported by the National Science Foundation under Grant PHY-0355005, and by 
the Department of Energy under Grant DE-FG02-90ER40542.

\end{document}